\documentclass{IEEEtran}
\usepackage{cite}
\usepackage{amsmath,amssymb,amsfonts}
\usepackage{algorithmic}
\usepackage{graphicx}
\usepackage{textcomp}
\def\BibTeX{{\rm B\kern-.05em{\sc i\kern-.025em b}\kern-.08em
    T\kern-.1667em\lower.7ex\hbox{E}\kern-.125emX}}
\begin{document}
\title{ Lambda/6 Suspended Patch Antenna  }

\author{Luca Giangrande
\thanks{Manuscript received October 15, 2009. This work was supported by the STW foundation (PLEISTER project, code: 10060).}
\thanks{
L. Giangrande is with CERN , Esplanade de Particules 1, Meyrin, Switzerland (e-mail: luca.giangrande@cern.ch). }}

\maketitle

\begin{abstract}
This work introduces a novel, compact antenna design based on a $\lambda/6$ suspended patch configuration that is particularly suited for small-size wireless sensor nodes adopting the worldwide available $2.45\,$GHz industrial, scientific and medical (ISM) band. The proposed design meets key requirements such as compactness, omnidirectionality in the azimuth plane, robust source matching over a designated bandwidth, interference immunity, and low costs. The design is an evolution of the conventional planar inverted f-antenna (PIFA). A minimal ground plane enables the shielding of the underlying circuitry as well as the reduction of the antenna's effective dimensions below one-half wavelength, when combined with a specific geometrical arrangement of the vertical conducting rods.

The simulation results, performed with a cost-effective FR4 substrate model, demonstrate a resonance at 2.45\,GHz with a return loss of –32.5\,dB and a bandwidth of 50\,MHz relative to the –10\,dB level. With a mechanical footprint of only $20 \times 20\, \text{mm}^2$, this antenna design constitutes an attractive candidate for integration into compact and densely populated wireless sensor networks.

\end{abstract}


\section{Introduction}

\begin{figure}
    \centering
    \includegraphics[width=0.4\textwidth]{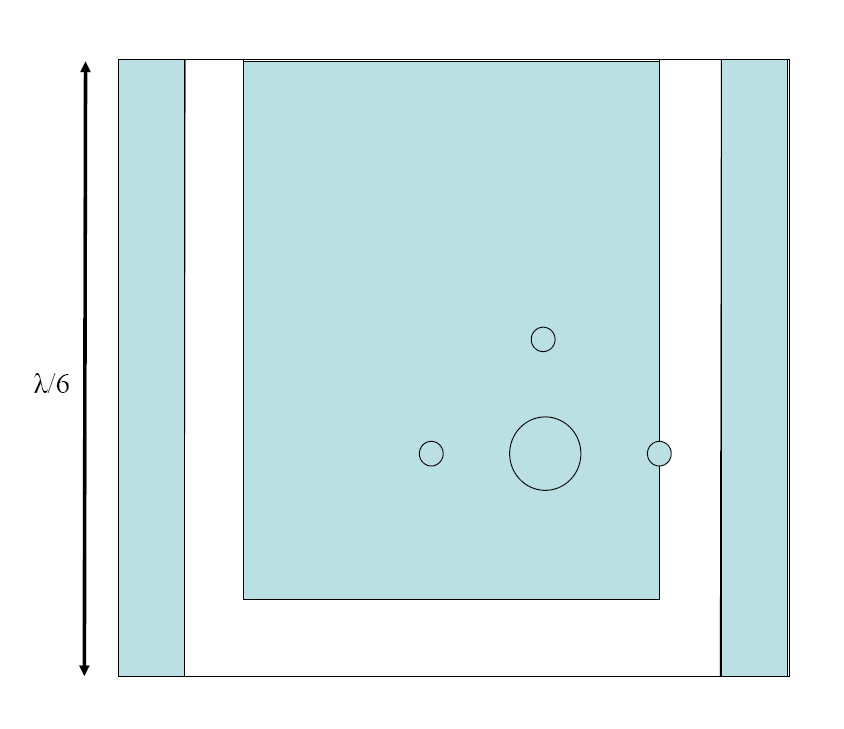}
    \caption{Top view of the PCB layout. $\lambda$ is the free space wavelength at the resonance frequency (2.45\,GHz). The bottom ground plane has the same size of the top structure bit it is not visible in the figure.}
    \label{fig:pcb}
\end{figure}

Wireless sensor networks are gaining interest in the scientific community due to their flexibility. In fact, depending on the type of sensors equipped, they can target different applications such as environment monitoring, biomedical diagnoses, bridge elasticity, goods tracking, etc.
The feasibility of wireless sensor networks has been made possible through technological progresses in achieving high scale of integration and reduced power consumption for both electronic circuitry and sensing devices. The high scale of integration allows equipping the nodes with high processing capabilities within a reduced surface area, ranging from several square millimeters to few square centimeters. 
A wireless sensor node is equipped with sensing, communicating and processing/storing capabilities. 
The intelligence of the network is created by distributed computing based on tight interaction between nodes. To make this interaction effective, the communication front–end is a critical element. The antenna is a part of the communication front–end and its purpose is to optimize the energy exchange between the transmitter and the receiver. If the problem is addressed from the standpoint of “Transmission Line Theory”, the optimization is achieved through load matching and the antenna itself can be used to match the transceiver impedance to the communication medium impedance. Owing to reciprocity, if the antenna performs a matching between the transmitting radio and the medium, it will also produce a match between the medium and the receiver radio, provided that the transmitter and receiver electronics present the same impedance. This property implies that an antenna optimized for transmission performs as good when employed for reception.
A key requirement for a radiator in compact wireless sensor nodes is minimal volume occupation while maintaining good efficiency. 
In the applications where the sensor nodes are installed on a surface, the ideal radiation pattern is hemispherical so that any neighbor can be successfully reached regardless to its angle of sight. To bind the radiation in the upper half plane and to reciprocally shield the antenna and the underlying electronic components of the node itself (sensors, computing and power management circuits, battery, etc.), a ground plane is placed under the radiating part of the antenna.  
Therefore the antenna is indented to be stacked on the top part of the wireless node. The higher elevation from the surface on which the node is placed makes communication more effective by improving the radiation mechanism. 
The antenna proposed is compatible with any dual-layer PCB process and suitable to be realized on a variety of materials; in this work the cheapest and most used for standard electronics substrate (FR4) has been selected.
The proposed design results in a more compact layout compared to the standard types of antennas with similar properties. For instance, a square patch antenna has a minimum size of $\lambda/2$, a planar inverted F antenna (PIFA) needs a ground plane at least $\lambda/3$ long whereas its thickness plus the length must equal $\lambda/4$ \cite{b1}. The antenna presented in this work (Figure \ref{fig:pcb}) is an evolution of a square patch antenna such as \cite{b2} but with a quite more compact layout: length of $\lambda/6$ and a thickness of $\sim\lambda/40$. 

\section{Simulation Results}
\subsection{Interaction with the transceiver}
\begin{figure}
    \centering
    \includegraphics[width=0.45\textwidth]{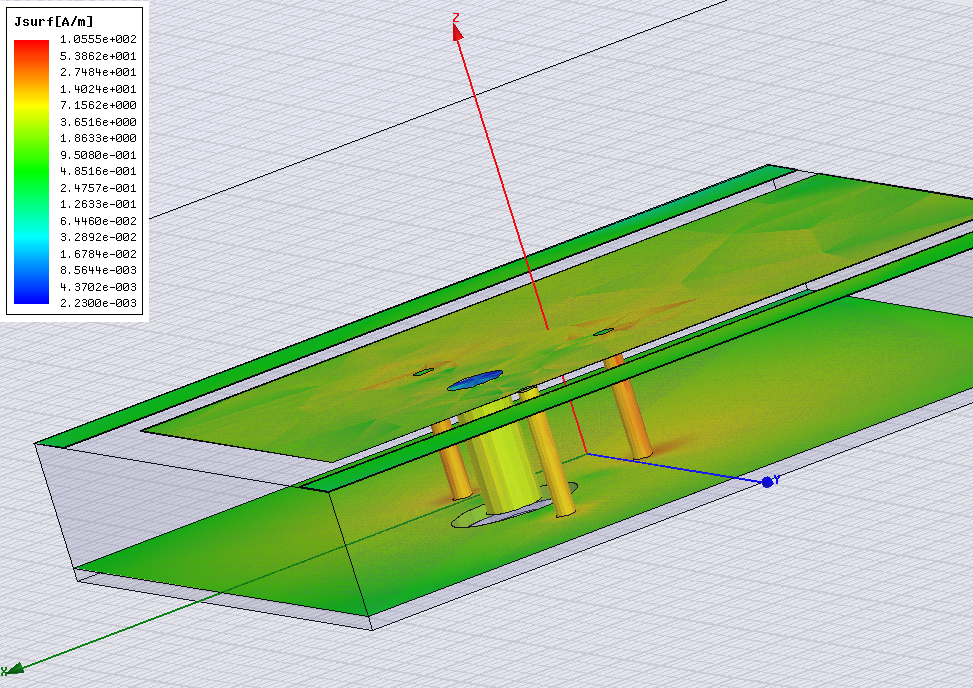}
    \caption{Finite elements simulation of the superficial current density distribution $[A/m]$ at resonance (2.45\,GHz). The highest density is reached on the vertical conducting rods. The depicted coordinate system is  is identical to that used in all the simulations presented in this article.}
    \label{fig:mech_3d}
\end{figure}

According to \cite{b3} the proposed radiator falls in the category of suspended plate antennas, having a thickness of 0.026 times the wavelength. The antenna's physical structure is a stack made of a dielectric substrate with two metal layers (Figure \ref{fig:mech_3d}) where the bottom layer is connected to the ground of the feed and the top layer is connected to the feed by a conducting pin going through the substrate. The top plate is then connected to the bottom one via three grounding pins.  Two metal strips are then placed near the top plate to adjust the capacitive part of the antenna impedance. To evaluate the antenna performances a commonly used and cheap PCB material (FR4) has been chosen as substrate, whereas copper is adopted as conductor. 
The position of the pins, shifted from the center of the ground plane, has a double influence in defining both the antenna impedance and its radiation pattern.

\begin{figure}
    \centering
    \includegraphics[width=0.45\textwidth]{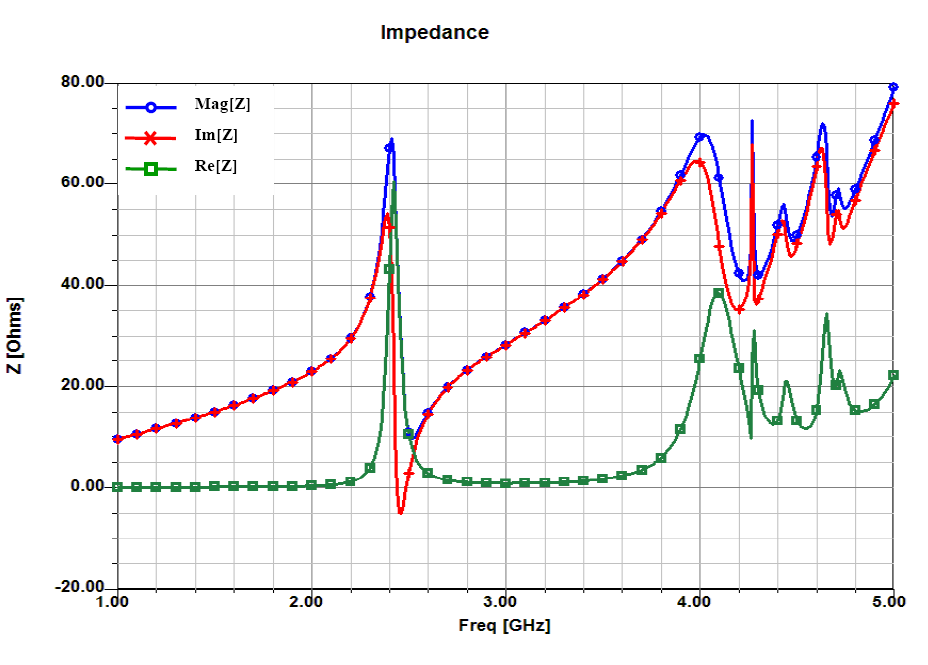}
    \caption{Simulation of the antenna's input impedance magnitude seen by the signal source as a function of the frequency. Blue: overall impedance. Green: real component. Red: imaginary component. The reactance exhibits a predominantly inductive behavior, as indicated by its positive slope with frequency. This characteristic behavior is associated with the current density through the vertical rods. At resonance, the capacitive component offsets the inductive component, resulting in a minimal net reactance. }
    \label{fig:impedance}
\end{figure}

The impedance of the radiator must be matched to the radio as to maximize the energy exchange. The return loss expresses the ratio of the reflected power ($P_r$) over the transmitted one ($P_t$) and can be directly calculated from the reflection coefficient:
\begin{equation}
  RL = 10 \log \left( \frac{P_r}{P_i} \right) = 10 \log \left( |\Gamma|^2 \right)
  \label{eq:reflection_loss}
\end{equation}
The return loss is found to be a function of the frequency since the reflection coefficient is defined as:
\begin{equation}
\Gamma(f) = \frac{Z_a(f) - Z_s(f)}{Z_a(f) + Z_s(f)}
\label{eq:reflection_coefficient}
\end{equation}
where $Z_a$ and $Z_s$ are respectively the antenna and source impedance. Typically, the feeding source exhibits an output impedance that is predominantly real-valued and remains relatively stable across the operational frequency band, with negligible reactive contribution.
\begin{figure}
    \centering
    \includegraphics[width=0.45\textwidth]{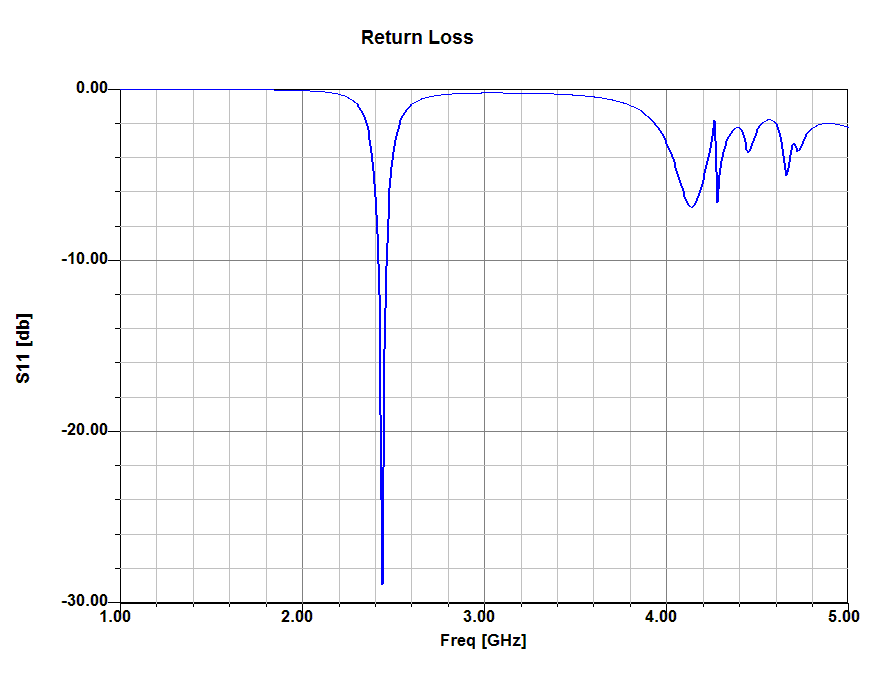}
    \caption{Simulation of the Return Loss as function of the frequency observed with a $50\,\Omega$ signal source. S11 refers to the reflected power at the input of a two-ports network.}
    \label{fig:return_loss}
\end{figure} 
This antenna shows a return loss (Figure \ref{fig:return_loss}) of –32.5\,dB at the frequency of 2.45\,GHz. The antenna bandwidth can be defined with respect to the 10\,dB attenuation of the return loss giving a value of 50\,MHz (from 2.42 to 2.47\,GHz). 
The resonance frequency is defined as the point at which the imaginary part of the impedance is null.  To achieve a good matching, the real part of the antenna impedance must be as close as possible to the source impedance. The dependence of the impedance on the frequency (Figure \ref{fig:impedance}) suggests that the antenna can be modeled with an equivalent parallel RLC circuit, in which the capacitance is related to the metal plates, whereas the pins are responsible for the resistance and inductance. 
\subsection{Interaction with the medium}
The emitted radiation pattern depends on how the current density is distributed over the antenna's conducting surfaces, previously shown in Figure \ref{fig:mech_3d}. 

Unlike a conventional patch antenna that radiates from its top conductor, this design uses the vertical feeding rods as the primary radiating elements. In fact, to their limited cross-section, the rods sustain current densities far exceeding those in the metal plates, making them the primary source for radiation. Beyond guiding current flow paths, the metal plates principally act as capacitive elements that enable a dramatic shortening of the antenna's electrical length.

\begin{figure}
    \centering
    \includegraphics[width=0.45\textwidth]{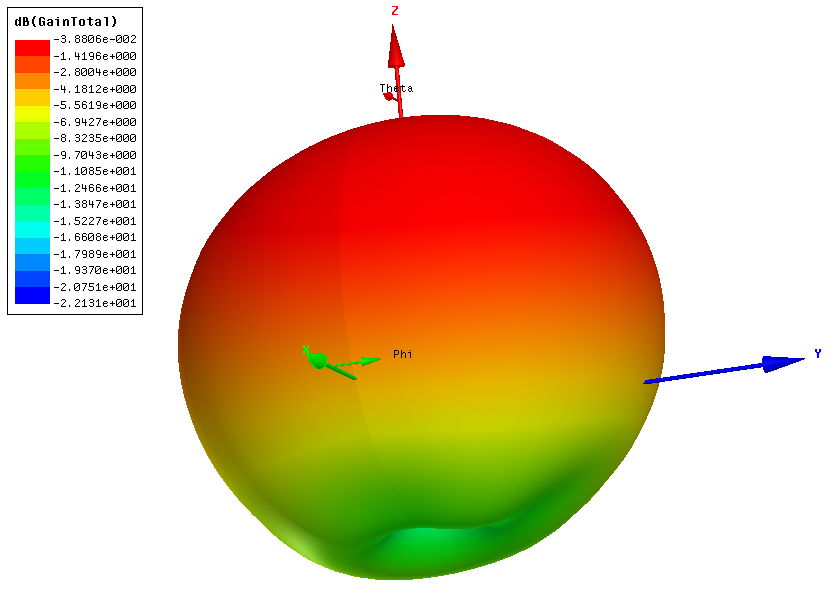}
    \caption{Side view of the simulated tridimensional radiation pattern at resonance ($2.45\,$GHz). The azimuth $\varphi$ and zenith $\theta$ angles are shown as well. The desired hemispherical configuration is obtained for positive z, whereas the radiation along negative z is minimized.}
    \label{fig:rad_3D}
\end{figure}

The radiation pattern characterizes both the spatial distribution of transmitted power (shown in Figure \ref{fig:rad_3D}) and, by reciprocity principle, the angular sensitivity for the received power. 

\begin{figure}
    \centering
    \includegraphics[width=0.45\textwidth]{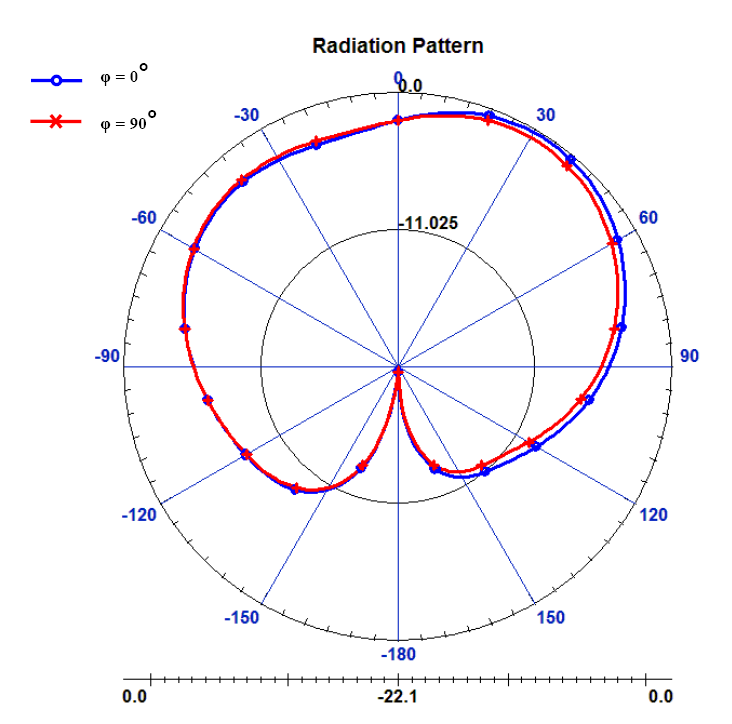}
    \caption{Polar representation of the radiation patterns in the elevation (or zenith) planes as a function of the zenith angle. Blue: XZ plane, $\varphi = 0^{\circ}$. Red: YZ plane, $\varphi = 90^{\circ}$. Radial direction represents the gain in dB.}
    \label{fig:rad_polar}
\end{figure}

\begin{figure}
    \centering
    \includegraphics[width=0.45\textwidth]{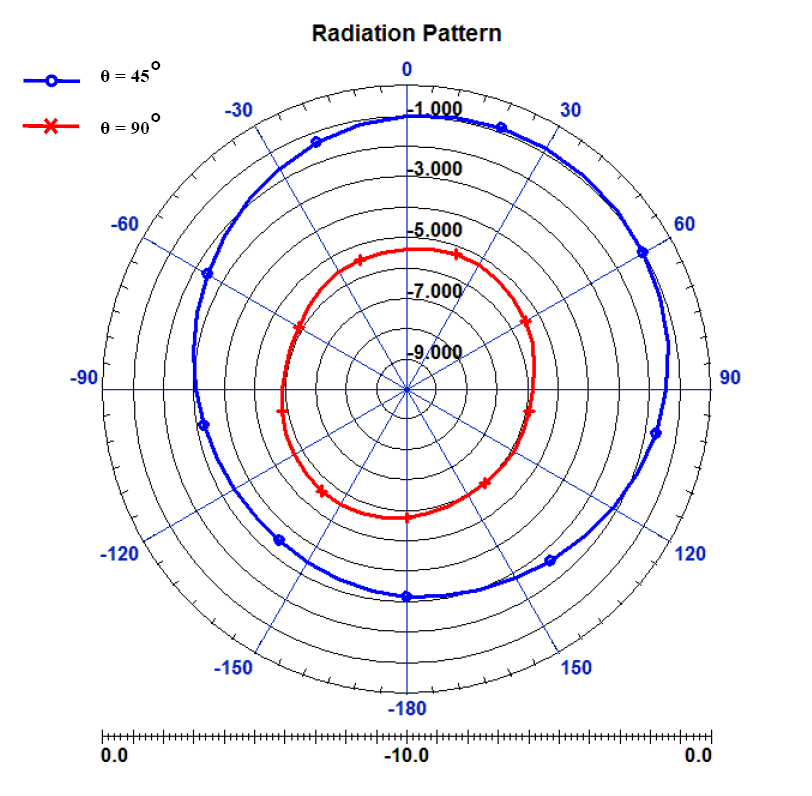}
    \caption{Polar representation of the radiation patterns as a function of the azimuth angle. Red: along the azimuth (XY) plane, $\theta = 90^{\circ}$. Blue along the cone at the oblique direction $\theta = 45^{\circ}$. Radial direction represents the gain in dB.}
    \label{fig:rad_polar2}
\end{figure}

The two-dimensional radiation pattern cuts in the elevation (XZ) and azimuth (XY) planes are presented as polar diagrams in Figures \ref{fig:rad_polar} and \ref{fig:rad_polar2}, respectively. The strongest omnidirectional characteristic is observed along the azimuth plane.

\begin{figure}
    \centering
    \includegraphics[width=0.45\textwidth]{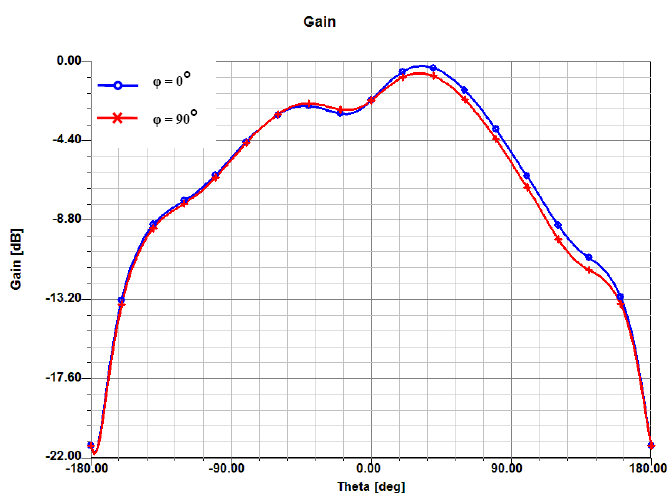}
    \caption{Representation of the antenna's gain, in dB, as a linear function (rectangular coordinates) of the zenith angle. Blue: XZ plane, $\varphi = 0^{\circ}$. Red: YZ plane, $\varphi = 90^{\circ}$. In both cases the maximum radiation is observed under a zenith of $\theta = 30^{\circ}$. }
    \label{fig:gain_theta}
\end{figure}

Antenna gain is a key performance metric that measures both directivity and efficiency, reflecting how effectively the antenna focuses power in particular directions. Figure \ref{fig:gain_theta} plots the simulated gain versus zenith angle, revealing that half of the radiated power is concentrated within a primary lobe spanning $–60^{\circ}< \theta <+70^{\circ}$. This corresponds to a half-power beamwidth (HPBW) of $130 ^{\circ}$.
  
The peak gain of –0.02\,dB occurs at an azimuth angle of $\varphi = 45^{\circ}$ and zenith of $\theta = 30^{\circ}$. Employing a low-loss substrate could significantly enhance the gain, achieving a maximum of 3.5\,dB. The gain value critically impacts the link budget by directly influencing the maximum achievable transmission range.

The antenna exhibits a radiation efficiency of 41\%, significantly limited by the substrate's high energy absorption. This is attributed to the material's poor RF characteristics (namely a typical loss tangent of 0.02 for FR4 substrates). Radiation efficiency, defined as the ratio of total radiated power to input power at the antenna port, can be decomposed into three multiplicative factors detailed in:
\begin{equation}
\varepsilon = \left( 1 - |\Gamma|^2 \right) \varepsilon_c \varepsilon_d
\label{eq:example}
\end{equation}
where $(1-|\Gamma|^2)$ represents the matching efficiency, $\epsilon_c$ the conduction efficiency and $\epsilon_d$ the dielectric efficiency. 
Therefore, total efficiency can be enhanced by optimizing any of the three contributing factors: improving impedance matching with the source, reducing conductor losses, or minimizing dielectric losses. While metallic losses, due to finite conductivity, have negligible impact on gain, dielectric losses play a dominant role in efficiency reduction. Substituting the substrate with a low-loss ceramic material—exhibiting a loss tangent ten times lower than FR4—is expected to improve efficiency up to 86\%.

\section{Conclusion}
 
A novel, $\lambda/6$ suspended patch antenna has been presented as a viable solution for compact wireless sensor nodes. The antenna’s design achieves a remarkably reduced footprint—only $20 \times 20\,\text{mm}^2$ by evolving the traditional square patch structure and integrating a grounded metal shield to isolate and protect the radiator from circuit interference. Simulation results confirm that the antenna resonates at 2.45 GHz with an excellent return loss of –32.5\,dB and a 50\,MHz bandwidth at the –10\,dB level, while also delivering a nearly omnidirectional radiation pattern in the azimuth plane with a half-power beamwidth of $130^\circ$. Although the current implementation on an FR4 substrate yields a radiation efficiency of 41\%, the design framework offers ample opportunities for further enhancements, such as substituting the substrate with a low-loss ceramic material to potentially boost efficiency up to 86\%.

In summary, the proposed antenna design not only meets but exceeds the typical constraints imposed on wireless sensor node communication modules in terms of size, cost, and performance. Its compact form factor, coupled with robust impedance matching and favorable radiation characteristics, makes it an excellent candidate for next-generation wireless sensor networks. Future work will explore material optimizations and integration strategies that further refine the performance of this design, paving the way for widespread adoption in diverse sensor network applications.

\end{document}